\documentclass[pdflatex,sn-mathphys-num]{sn-jnl}

\usepackage{graphicx}%
\usepackage{multirow}%
\usepackage{amsmath,amssymb,amsfonts}%
\usepackage{amsthm}%
\usepackage{mathrsfs}%
\usepackage[title]{appendix}%
\usepackage{xcolor}%
\usepackage{textcomp}%
\usepackage{manyfoot}%
\usepackage{booktabs}%
\usepackage{algorithm}%
\usepackage{algorithmicx}%
\usepackage{algpseudocode}%
\usepackage{listings}%

\theoremstyle{thmstyleone}%
\theoremstyle{thmstyletwo}%

\theoremstyle{thmstylethree}%

\begin{document}

\title[Black Holes, Gravitational Waves and Space-Time Singularities (Lema{\^\i}tre Conference 2024) -- Editorial]{Black Holes, Gravitational Waves and Space-Time Singularities (Lema{\^\i}tre Conference 2024) -- Editorial}







\author[1,2]{\fnm{Massimo} \sur{Bianchi} \orcid{https://orcid.org/0000-0002-7591-3870}}

\author[3,4,5]{\fnm{Sergio Luigi} \sur{Cacciatori}\orcid{https://orcid.org/0000-0002-4167-9123} }

\author[6,7,8]{\fnm{Matteo} \sur{Galaverni} \orcid{https://orcid.org/0000-0002-5247-9733}}

\author*[6,9]{\fnm{Gabriele} \sur{Gionti, S.J.}\orcid{https://orcid.org/0000-0002-0424-0648}}\email{ggionti@specola.va}

\author[10,11]{\fnm{Fabio} \sur{Scardigli}\orcid{https://orcid.org/0000-0002-9415-3256}}

\affil[1]{\orgdiv{Dipartimento di Fisica}, \orgname{Universit\`a di Roma `Tor Vergata'}, \orgaddress{\street{Via della Ricerca Scientifica 1}, \city{Roma}, \postcode{00133}, \country{Italy}}}

\affil[2]{\orgname{INFN, Sezione di Roma Tor Vergata}, \orgaddress{\street{Via della Ricerca Scientifica 1}, \city{Roma}, \postcode{00133}, \country{Italy}}}

\affil[3]{\orgdiv{DiSAT}, \orgname{Universit\`a dell'Insubria}, \orgaddress{\street{Via Valleggio 11}, \city{Como}, \postcode{22100}, \country{Italy}}}

\affil[4]{\orgname{INFN, Sezione di Milano}, \orgaddress{\street{ Via Celoria 16}, \city{Milano}, \postcode{20133}, \country{Italy}}}

\affil[5]{\orgdiv{Como Lake centre for AstroPhysics (CLAP), DiSAT}, \orgname{Universit\`a dell'Insubria}, \orgaddress{\street{ Via Valleggio 11}, \city{Como}, \postcode{22100}, \country{Italy}}}

\affil[6]{\orgname{Vatican Observatory}, \orgaddress{\city{Vatican City}, \postcode{00120}, \country{Vatican City State}}}

\affil[7]{\orgname{INAF/OAS Bologna}, \orgaddress{\street{via Gobetti 101}, \city{Bologna}, \postcode{40129}, \country{Italy}}}

\affil[8]{\orgname{INFN, Sezione di Bologna}, \orgaddress{\street{via Irenerio 46}, \city{Bologna}, \postcode{40126}, \country{Italy}}}

\affil[9]{\orgdiv{Laboratori Nazionali di Frascati}, \orgname{INFN}, \orgaddress{\street{Via E. Fermi 40}, \city{Frascati}, \postcode{00044}, \country{Italy}}}

\affil[10]{\orgdiv{Dipartimento di Matematica}, \orgname{Politecnico di Milano}, \orgaddress{\street{Piazza Leonardo da Vinci 32}, \city{Milano}, \postcode{20133}, \country{Italy}}}

\affil[11]{\orgdiv{Lorentz Institute for Theoretical Physics}, \orgname{Leiden University}, \orgaddress{\street{P.O. Box 9506}, \city{Leiden}, \country{The Netherlands}}}




\maketitle

\section{Introduction}\label{sec1}

This editorial introduces the topical collection arising from the Lema{\^\i}tre Conference 2024\footnote{Conference website: \url{https://indico.global/event/13754/}} 
held at the Vatican Observatory in Castel Gandolfo, 17--21 June 2024.  The 19 contributions collected here address the most pressing open problems at the interface of cosmology, gravitation, and quantum theory: the Hubble tension and the future of $\Lambda$CDM in light of recent DESI data; inflationary cosmology and dark-energy model-building in string theory and supergravity, together with the Swampland constraints that bound them; the observer-dependence of the quantum-cosmological wave function; the removal, or consistent crossing, of spacetime singularities; primordial black holes and their gravitational-wave signatures; the search for a background-independent formulation of quantum gravity and a first-principles account of horizon thermodynamics; and the foundational status of the quantum-to-classical transition in gravitational contexts. A proposal for lunar-based cosmological observations, and a historical reconstruction of the genesis of Lema{\^\i}tre's primeval-atom hypothesis, complete the collection.

The legacy of Georges Lema{\^\i}tre --- whose 1927 derivation of the recession of the nebulae\footnote{Lema{\^\i}tre, G.: Republication of `A homogeneous universe of constant mass and increasing radius accounting for the radial velocity of extra-galactic nebulae'. Gen. Relativ. Gravit. {\bf 45}, 1635--1646 (2013), \url{https://doi.org/10.1007/s10714-013-1548-3}.} and 1931   Hypothesis of the primeval atom\footnote{Lema{\^\i}tre, G.: The expanding universe. Mon. Not. R. Astron. Soc. {\bf 91}, 490--501 (1931), \url{https://doi.org/10.1093/mnras/91.5.490}.} laid the conceptual foundations of the Big Bang paradigm --- provides both a historical anchor and an enduring source of inspiration for the research presented in this collection. 
Lema{\^\i}tre's own contributions to the themes discussed here were not incidental: it was he who first recognized, 
in his work on the static de Sitter metric in the 1920s\footnote{Lema{\^\i}tre, G.: Note on de Sitter's Universe. J. Math. Phys. (MIT) {\bf 4}, 188--192 (1925), \url{https://doi.org/10.1002/sapm192541188}.}, 
that an apparent singularity could be a mere artifact of coordinates rather than a true breakdown of the geometry, and it was he who coined the term ``horizon'' for such a locus. In 1933, in \textit{``L'Univers en expansion''} \footnote{Lema{\^\i}tre, G.: Republication of `The Expanding Universe'. Gen. Relativ. Gravit. {\bf 29}, 641--680 (1997), \url{https://doi.org/10.1023/A:1018855621348}.},
he extended this insight to the Schwarzschild solution, anticipating by a quarter of a century the modern understanding of the horizon as a coordinate rather than a physical singularity, and, in the same paper, introduced the inhomogeneous dust solution that today bears his name alongside those of Tolman and Bondi.

The Lema{\^\i}tre Conference 2024
was the second in a series of workshops dedicated to this legacy, following the inaugural meeting of 2017\footnote{See the special issue in Found. Phys. {\bf 48}(10) (2018): \url{https://link.springer.com/collections/abchcehbfc}.}. The combination of formal talks, extended daily discussion, and the uniquely welcoming and intellectually stimulating atmosphere of the Vatican Observatory, proved exceptionally productive. Specialists in cutting-edge observation, theoretical physicists, and historians of science were able to develop a shared language and to identify the most pressing open challenges in astrophysics, cosmology, and quantum gravity. The present topical collection gathers contributions from the invited speakers, offering a representative survey of the current state of the field.

\section{Cosmology and the Hubble tension}\label{sec2}
One of the central topics of the conference was the so-called Hubble tension --- the statistically significant discrepancy between independent measurements of the present-day expansion rate of the universe. Determinations based on the distance-ladder method, anchored by Type~Ia supernovae calibrated with Cepheid variables, yield values of the Hubble constant that differ substantially from those inferred from the temperature anisotropies of the Cosmic Microwave Background (CMB) within the standard  $\Lambda$CDM framework. At present, this tension --- now standing at roughly 5$\sigma$ --- remains unresolved, and may reflect uncharacterised systematics in one or both measurement chains, or a genuine breakdown of the standard cosmological model.

Michael S. Turner offered a critical status report on $\Lambda$CDM, framed explicitly as a sequel to his talk at the 2017 Lema{\^\i}tre meeting. While the standard model continues to fit the overwhelming majority of data, Turner focused on the DESI Collaboration's evidence (DR1 and DR2) for a time-evolving dark-energy equation of state in the $w_0$--$w_a$ parametrisation, whose most provocative feature is a dark-energy density sharply peaked around $z \sim 0.5$ rather than the constant value predicted by a cosmological constant. 

Joseph Silk's contribution likewise opened from the observation that ``modern cosmology effectively began with Georges Lema{\^\i}tre in 1927,'' and is discussed further, together with its proposal for a new observational frontier.
His proposed solution is a sustained scientific presence on the lunar far side, whose radio silence, seismic quietness, and lack of atmosphere make it uniquely suited to several genuinely guaranteed measurements
The discussions underlined, more broadly, the transformative impact that data from the James Webb Space Telescope has already had in testing the robustness of $\Lambda$CDM at high redshift, and the corresponding promise of a still more ambitious lunar observational frontier.

\section{Inflation, dark energy, and the string landscape}\label{sec3}
Inflationary cosmology remains a cornerstone of our understanding of the early universe, explaining the observed homogeneity, isotropy, and flatness of space and the origin of the primordial density fluctuations that seeded cosmic structure. Michele Cicoli presented recent progress on inflation and dark energy within type IIB string compactifications. On the inflationary side, his review centred on the Loop Blow-up Inflation scenario, in which a blow-up Kähler modulus with an approximate shift symmetry drives slow-roll inflation through a potential generated by string-loop corrections; the model yields sharp, falsifiable predictions --- a scalar spectral index in the narrow range $0.9757 \lesssim n_s \lesssim 0.9765$ and a tensor-to-scalar ratio $r \sim 2 \times 10^{-5}$ --- in excellent agreement with current CMB and BAO data. 
On dark energy, Cicoli surveyed the difficulty of realising quintessence in a UV-complete setting, presenting a two-axion hilltop model exploiting poly-instanton suppression as the most promising route to a phenomenologically viable, string-derived dynamical dark energy, while noting that de Sitter vacua remain comparatively easier to construct. Renata Kallosh and Andrei Linde reviewed the current status of inflationary cosmology from the perspective of supergravity, highlighting the predictive successes of attractor models alongside the open challenge of embedding inflation in a UV-complete framework.

The broader implications of quantum gravity for the space of viable low-energy theories were explored through the Swampland programme. Hirosi Ooguri reviewed Swampland-type constraints on effective theories of quantum gravity in asymptotically anti-de Sitter spacetimes, where such constraints can be rigorously tested via the AdS/CFT correspondence --- summarising, in particular, a proof (with D. Harlow) that any exact global symmetry in a bulk gravitational theory is incompatible with the consistency of the dual boundary CFT, and a further result establishing universal bounds on the exponential decay rate governing the Distance Conjecture in two-dimensional CFTs dual to AdS$_3$ gravity. Cumrun Vafa's contribution combined the Distance Conjecture, the associated species scale, the de Sitter Conjecture, and the TransPlanckian Censorship Conjecture (TCC) to derive increasingly sharp bounds on inflationary potentials, showing that imposing the TCC renders standard slow-roll inflation viable only in a strongly fine-tuned corner of parameter space with an essentially unobservable tensor-to-scalar ratio. Applied to the present epoch, the same reasoning implies that our universe --- if presently in a metastable de Sitter phase --- cannot remain so for much longer than of order two trillion years, a bound Vafa explicitly frames as string theory's answer to the question posed by the very title of Lema{\^\i}tre's 1927 paper, now viewed from the vantage of the universe's future rather than its origin.

\section{Quantum cosmology and the wave function of the universe}\label{sec4}
Thomas Hertog's contribution returned most directly to Lema{\^\i}tre's own 1931 Nature letter on the primeval atom, reading it as an early and remarkably prescient statement that the origin of the universe should be a proper object of physical, rather than merely metaphysical, inquiry. Tracing a conceptual line from Lema{\^\i}tre's primeval quantum through the Hartle--Hawking no-boundary wave function to the modern ``top-down'' reformulation of quantum cosmology, Hertog addressed a long-standing embarrassment of the no-boundary proposal: taken at face value, it overwhelmingly favours nearly empty histories incompatible with the existence of observers. He showed that once an observer is treated as a genuine quantum subsystem within the theory --- modelled concretely via the information content of a CMB temperature map --- the resulting conditional probability distribution can undergo a Page-like transition, in which the dominant saddle point shifts abruptly from a low-inflation history to one beginning deep in the eternal-inflation regime for sufficiently detailed observational situations. The past, on this view, is contingent on the question being asked of the wave function --- a striking modern echo of Lema{\^\i}tre's own insistence, at the 1958 Solvay Council, that any information on the state of matter must be inferred from the condition that the actual universe has been able to evolve from it.

\section{Singularities, black holes, and gravitational waves}\label{sec5}
The nature of spacetime singularities --- whether at the Big Bang or in the deep interior of black holes --- remains one of the most profound unresolved problems in theoretical physics. General relativity predicts their inevitability, yet a complete theory of quantum gravity is widely expected to provide a regular, singularity-free description. Two contributions took Lema{\^\i}tre's own 1933 dust model as their explicit point of departure. Claus Kiefer and Hamid Mohaddes, in their quantum treatment of Lema{\^\i}tre's model for gravitational collapse, asked what happens to its classical singularity under canonical quantization: working first with a thin null dust shell and then with the full Lema{\^\i}tre--Tolman--Bondi cloud, reduced shell by shell to a self-adjoint Hamiltonian for the outermost layer, they constructed exact, normalizable wave-packet solutions whose unitary evolution forces the collapsing packet to bounce at a minimal radius and re-expand, rather than terminate --- a picture that persists in the homogeneous Oppenheimer--Snyder limit under affine coherent-state quantization, though the authors remain open about whether the resulting bounce timescale is compatible with observation and whether the method extends beyond spherical symmetry.

Alexander Kamenshchik revisited the problem of singularity crossing, developed since the first Lema{\^\i}tre Conference of 2017, focusing on the transition between the Jordan and Einstein conformal frames: a Big Bang--Big Crunch singularity in one frame can correspond to a perfectly regular geometry in the other, allowing the crossing to be described unambiguously. First worked out for isotropic Friedmann--Lema{\^\i}tre universes, the idea has since been extended to anisotropic Bianchi-I and Kantowski--Sachs cosmologies and to a covariant field-space formalism characterizing which singularities are removable by field reparametrization; a parallel strand addresses quantum cosmology in the Wheeler--DeWitt sense, showing that for soft future singularities such as the ``Big Brake'' the wave function can vanish at the singularity while the correctly normalised probability density does not --- a position Kamenshchik defends, echoing Charles Misner's 1969 remark, as making the coexistence with singularities itself a legitimate research programme.
Gabriele Veneziano presented recent progress on the central open problem of the Pre-Big-Bang scenario he proposed with Maurizio Gasperini over three decades ago: whether the singularity separating the inflationary pre-bang branch from the decelerating post-bang branch can be tamed by higher-order $\alpha'$ corrections consistent with the $O(d,d)$ duality symmetry of classical string cosmology. Building on the all-order reformulation of Hohm and Zwiebach, a Hamiltonian/Routhian approach reduces the existence of regular, bouncing solutions to a simple analytic criterion, yielding explicit ``clockwise'' bounces and, with the addition of a non-perturbative dilaton potential, late-time attractors of Minkowski, metastable-vacuum, or de Sitter type, together with a new isotropization mechanism in anisotropic extensions.

At a more fundamental level, Roberto Casadio questioned the common assumption that quantum gravity is relevant only at the Planck length, arguing that this conflates the Compton length governing scattering with the very different scales governing bound states. Proposing that quantum effects become important for any self-gravitating system whose compactness approaches unity, he built a many-body ground state for a dust ball from a hierarchy of quantised shells, each obeying a hydrogen-atom-like radial equation, finding a core radius of order the gravitational radius, with the ground-state occupation number reproducing the Bekenstein area scaling; the resulting interior has finite tidal forces and no inner Cauchy horizon, replacing the point singularity with what Casadio calls an integrable singularity, a picture recovered independently from coherent states of an auxiliary scalar field and extended to slowly rotating geometries. Misao Sasaki reviewed the formation of primordial black holes from rare, large-amplitude curvature perturbations on CMB-unconstrained inflationary scales, with non-minimally coupled curvaton models capable of producing Primordial Black Hole dark matter in the asteroid-mass window ($10^{18}$--$10^{22}$\,g) together with a scalar-induced gravitational-wave background within reach of forthcoming detectors such as LISA. Gia Dvali proposed a microscopic, string-theoretic account of de Sitter horizon entropy via open--closed string duality, showing, in a D9--anti-D9 brane construction in the 't~Hooft limit, that at a critical coupling the species entropy of the open-string degrees of freedom exactly reproduces the closed-string Gibbons--Hawking entropy, with implications for the ``memory burden'' effect and the quantum-breaking time of de Sitter space.

\section{Quantum gravity, horizons, and emergent spacetime}\label{sec6}
A major focus of the conference was the search for a consistent theory of quantum gravity, and, in particular, for a first-principles account of the horizon thermodynamics that Lema{\^\i}tre himself was the first to name. Edward Witten presented progress towards a background-independent algebraic formulation of quantum gravity, constructing an algebra of observables --- fields gravitationally dressed to the worldline of an observer with bounded-below energy --- defined without reference to any particular background spacetime and becoming background-dependent only once a Hilbert-space representation is chosen. Specialised to a geodesic observer in empty de Sitter space, the algebra acquires a genuine trace, and the thermal Bunch--Davies state of maximum entropy reproduces, via its vanishing relative entropy, Bousso's intuition that the late-time, empty static patch is the entropically preferred state; Witten closed by conjecturing that the Hartle--Hawking no-boundary state may play an analogous universal role beyond de Sitter space, connecting his construction to the observer-dependent quantum cosmology discussed by Hertog in Sect.~\ref{sec4}.

Raphael Bousso and Sami Kaya extended the notion of a generalised entanglement wedge, or ``hologram,'' from AdS/CFT boundary regions to arbitrary gravitating regions, via the fundamental complement of a bulk wedge --- the smallest wedge containing all causal curves of infinite proper duration lying entirely in its spacelike complement. This yields a full complementarity theorem for holograms and, of particular cosmological interest, shows that any spacetime containing a Big Bang or Big Crunch is ``trivially reconstructible'': the entanglement wedge of any wedge is the entire universe, an information-theoretic counterpart to Lema{\^\i}tre's own intuition that a genuine cosmological beginning renders any pre-existence of the universe causally inaccessible --- while de Sitter space escapes this trivial reconstructibility whenever the fundamental complement fails to vanish. Batoul Banihashemi and Ted Jacobson took up the Gibbons--Hawking derivation of the Bekenstein--Hawking entropy $A/4G$ from the Euclidean gravitational path integral, arguing that this near-half-century-old result rests on shaky foundations --- the Euclidean Einstein--Hilbert action is unbounded below, and the correct integration contour is unknown. Reviewing recent attempts at a first-principles derivation, they showed how a Lorentzian version of the Gauss--Bonnet theorem, combined with a simplicial (Regge calculus) treatment of the horizon's deficit angle, can reproduce the Bekenstein--Hawking result with a specific, physically motivated choice of sign.

\section{Foundations of quantum theory and semiclassical gravity}\label{sec7}
The conference devoted considerable attention to foundational issues in quantum mechanics as they arise in cosmological and gravitational settings. Rosa-Laura Lechuga-Solis and Daniel Sudarsky critically examined the routine identification, in inflationary cosmology, of quantum uncertainties with genuine stochastic fluctuations, arguing that this conflation obscures the unresolved quantum measurement problem in a setting devoid of external observers; adopting a semiclassical self-consistent configuration framework supplemented by spontaneous collapse (CSL-type) dynamics, they derived modified power spectra with a substantially suppressed tensor-to-scalar signal, and a mechanism by which the eternal-inflation problem may be circumvented. Lajos Di\'osi's stochastic formulation of semiclassical gravity is designed to avoid the conceptual difficulties associated with macroscopic Schr\"odinger-cat superpositions in gravitating systems; building on the non-relativistic Di\'osi--Penrose theory of gravity-induced spontaneous collapse, he formulated a ``healthier'' semiclassical dynamics based on spontaneous quantum monitoring and feedback, restoring linearity and reducing, in the Newtonian limit, to a modified Schrödinger--Newton equation free of the Born-rule violations of the conventional formulation --- while concluding that a relativistic extension of this and related ``post-quantum gravity'' proposals remains obstructed chiefly by the absence of a consistent relativistic theory of continuous quantum monitoring. These contributions prompted lively debate, tracing back to the Einstein--Bohr controversy, about whether quantum mechanics provides a complete description of physical reality and about the extent to which the gravitational field may play a dynamical role in the measurement process.

\section{The historical and conceptual legacy of Lema{\^\i}tre}\label{sec9}
An important dimension of the conference was the sustained reflection on the historical and intellectual contributions of Monsignor Georges Lema{\^\i}tre. Dominique Lambert offered a historical and epistemological reconstruction of the genesis of Lema{\^\i}tre's 1931 primeval-atom hypothesis, tracing its roots to Lema{\^\i}tre's engagement with cosmic-ray physics (Millikan--Cameron), his 1930--1931 work on quantum theory (Heisenberg's uncertainty relations, Eddington's Dirac equation and Clifford algebra), and his response to Eddington's philosophical rejection of a cosmic beginning. The paper distinguished the shifting ontological status the hypothesis held across Lema{\^\i}tre's career --- a logically prior quantum state preceding space-time, a physical ``giant nucleus,'' and an initial singularity --- and situated it within Lema{\^\i}tre's distinction between ``cosmogony,'' as a generative, non-formalised image, and rigorous mathematical cosmology, alongside his explicit theological separation of physical ``natural beginning'' from metaphysical creation. This historical perspective ran as a unifying thread throughout the conference, reminding participants that the open questions they were debating are not new, but have been at the heart of scientific and philosophical inquiry since Lema{\^\i}tre's own time.

\section{Conclusions}

The Lema{\^\i}tre Conference 2024 achieved its principal goal of fostering productive interaction between theory and observation across a wide range of sub-disciplines in cosmology and fundamental physics. The 19 contributions gathered in this topical collection reflect the richness and diversity of current research, as well as the many open questions that remain.

Progress on the Hubble tension will require both improved observational precision—across all rungs of the distance ladder and through new independent methods—and theoretical creativity in exploring extensions of and alternatives to $\Lambda$CDM. In particular, as discussed in the conference, is  the present late-time acceleration due to a true cosmological constant or something dynamical? The resolution of spacetime singularities awaits a formulation of a quantum theory of gravity. Based on the main conference talks, this quantum theory of gravity has to face the problem of the beginning of the universe, while the horizon thermodynamics addresses the searching of a fundamental theory able to explain the semi-classical and thermodynamical behaviour. 
While the interplay between inflation and string theory continues to raise deep questions about the very nature of physical law. 
At the foundational level, the relationship between quantum mechanics, measurement, and gravity remains one of the deepest open problems in theoretical physics.

The enduring example of Georges Lema{\^\i}tre --- a scientist who combined mathematical rigour, physical intuition, philosophical sophistication, and intellectual courage --- serves as an inspiration for all these endeavours. We hope that the contributions in this collection will stimulate further progress and encourage the mutual enrichment of ideas that is the hallmark of the best scientific conferences.

\section{Contents of the collection}
The papers collected in this topical issue are organised thematically as follows.

\subsection*{Cosmology and the Hubble tension}

\begin{itemize}
    \item 
  Turner, M.S.: $\Lambda$CDM: the path forward. Gen. Relativ. Gravit. {\bf 58}, 18 (2026). \href{https://doi.org/10.1007/s10714-026-03520-7}{[DOI]}
    \item 
  Silk, J.: The limits of cosmology. Gen. Relativ. Gravit. {\bf 57}, 127 (2025). \href{https://doi.org/10.1007/s10714-025-03450-w}{[DOI]}

\end{itemize}

\subsection*{Inflation, dark energy, and string theory}
\begin{itemize}
    \item 
  Cicoli, M.: Recent progress on inflation and dark energy from string theory. Gen. Relativ. Gravit. {\bf 58}, 33 (2026). \href{https://doi.org/10.1007/s10714-026-03538-x}{[DOI]}
    \item 
  Kallosh, R., Linde, A.: On the present status of inflationary cosmology. Gen. Relativ. Gravit. {\bf 57}, 135 (2025). \href{https://doi.org/10.1007/s10714-025-03470-6}{[DOI]}
    \item 
  Ooguri, H.: Constraints on quantum gravity. Gen. Relativ. Gravit. {\bf 57}, 123 (2025). \href{https://doi.org/10.1007/s10714-025-03455-5}{[DOI]}
    \item 
  Vafa, C.: On the origin and fate of our universe. Gen. Relativ. Gravit. {\bf 57}, 19 (2025). \href{https://doi.org/10.1007/s10714-025-03353-w}{[DOI]}

\end{itemize}

\subsection*{Quantum cosmology}
\begin{itemize}
    \item 
  Hertog, T.: A Page-like transition in quantum cosmology. Gen. Relativ. Gravit. {\bf 58}, 8 (2026). \href{https://doi.org/10.1007/s10714-025-03511-0}{[DOI]}

\end{itemize}

\subsection*{Singularities, black holes, and gravitational waves}
\begin{itemize}
    \item 
  Veneziano, G.: Recent progress in classical string cosmology. Gen. Relativ. Gravit. {\bf 57}, 151 (2025). \href{https://doi.org/10.1007/s10714-025-03482-2}{[DOI]}
    \item 
  Kamenshchik, A.: Again about singularity crossing in gravitation and cosmology. Gen. Relativ. Gravit. 56, 133 (2024). \href{https://doi.org/10.1007/s10714-024-03320-x}{[DOI]}
    \item 
  Sasaki, M.: Primordial black holes and gravitational waves from inflation. Gen. Relativ. Gravit. {\bf 57}, 82 (2025). \href{https://doi.org/10.1007/s10714-025-03412-2}{[DOI]}
    \item 
  Casadio, R.: The scale(s) of quantum gravity and integrable black holes. Gen. Relativ. Gravit. 56, 129 (2024). \href{https://doi.org/10.1007/s10714-024-03318-5}{[DOI]}
    \item 
  Dvali, G.: A string theoretic derivation of Gibbons-Hawking entropy. Gen. Relativ. Gravit. {\bf 57}, 118 (2025). \href{https://doi.org/10.1007/s10714-025-03446-6}{[DOI]}
    \item 
  Kiefer, C., Mohaddes, H.: Quantum theory of the Lema{\^\i}tre model for gravitational collapse. Gen. Relativ. Gravit. {\bf 57}, 26 (2025). \href{https://doi.org/10.1007/s10714-025-03349-6}{[DOI]}

\end{itemize}

\subsection*{Quantum gravity and the structure of spacetime}
\begin{itemize}
    \item 
  Witten, E.: A background independent algebra in quantum gravity. Gen. Relativ. Gravit. {\bf 57}, 17 (2025). \href{https://doi.org/10.1007/s10714-025-03360-x}{[DOI]}
    \item 
  Bousso, R., Kaya, S.: Fundamental complement of a gravitating region. Gen. Relativ. Gravit. {\bf 57}, 124 (2025). \href{https://doi.org/10.1007/s10714-025-03462-6}{[DOI]}
    \item 
  Banihashemi, B., Jacobson, T.: The enigmatic gravitational partition function. Gen. Relativ. Gravit. {\bf 57}, 43 (2025). \href{https://doi.org/10.1007/s10714-024-03347-0}{[DOI]}

\end{itemize}

\subsection*{Foundations of quantum theory and semiclassical gravity}
\begin{itemize}
    \item 
  Lechuga-Solis, RL., Sudarsky, D.: Addressing the so called quantum/classical ``divide'' in gravitational contexts, and its implications in cosmology. Gen. Relativ. Gravit. {\bf 57}, 86 (2025). \href{https://doi.org/10.1007/s10714-025-03420-2}{[DOI]}
    \item 
  Diósi, L.: A healthier stochastic semiclassical gravity: world without Schrödinger cats. Gen. Relativ. Gravit. {\bf 57}, 62 (2025). \href{https://doi.org/10.1007/s10714-025-03396-z}{[DOI]}

\end{itemize}

\subsection*{Historical perspectives}
\begin{itemize}
    \item 
  Lambert, D.: The origin(s) and meaning(s) of the primeval atom hypothesis: quantum physics meets Lema{\^\i}tre's cosmology. Gen. Relativ. Gravit. {\bf 57}, 105 (2025). \href{https://doi.org/10.1007/s10714-025-03422-0}{[DOI]}

\end{itemize}

\bmhead{Acknowledgements}

The conference was hosted at the Vatican Observatory in Castel Gandolfo and was co-sponsored by the Vatican Observatory and the Istituto Nazionale di Fisica Nucleare (INFN).

\section*{Declarations}

\bmhead{Conflict of interest} The authors declare no conflict of interest


\end{document}